\documentclass[preprintnumbers,showpacs,amsmath,amssymb,floatfix,9pt,prd,onecolumn,
superscriptaddress,nofootinbib]{revtex4}
\usepackage{latexsym}
\usepackage{epsfig}
\usepackage{amssymb}
\begin{document}

% Title

\title{Braneworld gravastars admitting conformal motion}

\author{Ayan Banerjee }
\email{ayan\_7575@yahoo.co.in} \affiliation{Department of
Mathematics, Jadavpur University, Kolkata 700 032, West Bengal,
India}
\author{Farook Rahaman}
\email{rahaman@iucaa.ernet.in
} \affiliation{Department of
Mathematics, Jadavpur University, Kolkata 700 032, West Bengal,
India}
\author{Sayeedul Islam}
\email{sayeedul.jumath@gmail.com} \affiliation{Department of
Mathematics, Jadavpur University, Kolkata 700 032, West Bengal,
India}
\author{Megan Govender}
\email{megandhreng@dut.ac.za} \affiliation{Department of Mathematics, Faculty of Applied Sciences,
Durban University of Technology, Durban 4000, South Africa}

\date{\today}

\begin{abstract}
In this paper, we propose the braneworld  gravastar configuration which is
alternative to braneworld black hole.
We study the Mazur and Mottola gravastar model within the context of Randall-Sundrum II type
braneworld scenario, based on the fact that our four dimensional  space-time is a three-brane,
embedded in a five dimensional bulk.  We present exact solutions of the modified field equations in each of the three regions making up the gravastar, namely, (I) the core, (II) the shell, and (III) the vacuum exterior.  The junction conditions at each interface are fulfilled and we further explore  interesting
physical properties such as length and energy and entropy of the spherical distribution.

\end{abstract}

\maketitle

\section{Introduction}
Black holes are regions in spacetime where gravity is so intense that even light cannot escape to the exterior.
These exotic objects have been the central focus of researchers over several decades.
The extraordinary physics of black holes has its origin in the highly unexpected properties of the
event horizon. The event horizon of a black hole is the boundary between its exterior and its interior;
which acts like a one-way membrane providing a strong connection between gravitation,
thermodynamics and quantum theory. This connection was first introduced by Hawking and Bekenstein
around 1974, when they looked at the nature of radiation emitted by black holes. It was soon
realized that this prediction created an information loss problem (the black hole information paradox) which has become an
important issue in quantum gravity and poses serious challenges to the foundations of theoretical physics. They
showed that black holes surrounded by quantum fields will emit small amounts of radiation,
causing them to shrink and eventually evaporating completely. These intrinsic problems associated with black hole horizons has led to a flurry of solutions attempting to remove these paradoxes. These semiclassical approaches generated by black hole horizons should be investigated because of the full
theory of quantum gravity is still unknown today.

Given the above, it has been suggested that alternative models of black holes, which do not involve
 horizons and could be stabilized under the exotic states of matter, should be studied.
Among the various models proposed and studied thus far, the gravitational vacuum star (gravastar)
has recently received widespread attention. The gravastar model was first proposed by Mazur and Mottola (MM)\cite{Mazur}.
In this model it is suggested that a gravitationally collapsing star would force spacetime
itself to undergo a phase transition that would prevent further collapse. The MM gravastar model consists of three regions:
a de Sitter geometry in the interior filled with constant positive (dark)
 energy density $\rho$ accompanied by isotropic negative pressure $\rho = -p >0$ which is connected
via three intermediate layers to an outer vacuum Schwarzschild solution ($p = \rho = 0$).  The intermediate relatively thin shell is composed of stiff matter ($p = \rho$). In order to achieve stability and compensate for discontinuities in the pressure profiles of the entire object, the MM gravastar requires two infinitesimally-thin shells endowed with surface densities $\sigma_\pm$ and surface tensions $\vartheta_\pm$. As pointed out by Cattoen et al. \cite{cattoen} this 5-layer construction can be reduced to a 3-layer model by effectively removing the use of thin shells. The gravastar is now composed of three regions:
(a) de-Sitter interior: $0 \leq r \leq r_1$, with equation of state $p = -\rho > 0$, (b) shell: $r_1 < r < r_2$ with equation of state $p = +\rho$ and (c) exterior: $r_2 < r$ with $p = \rho = 0$. Visser and Wiltshire \cite{Visser}
showed that the MM gravastar is dynamically stable against radial perturbations.
Subsequent work by Carter Chirenti \cite{Carter,Chirenti} and Rezzolla and De Benedictis \cite{DeBenedictis}
et al. have shown the MM gravastar is also stable against axial-perturbations. Investigations by Chan et al. \cite{Chan}
have shown that anisotropy of the interior fluid may affect the
formation of gravastar. They were able to show that formation of a gravastar is possible only when the tangential pressure is greater than
the radial pressure, at least in the neighborhood of the isotropic case. Numerous studies of gravastar models and their evolution can be found in \cite{Chan(2009),Rahaman,Horvat,Lobo,Turimov,Chirenti(2008)}, amongst others.

The Randall-Sundrum (RS) brane-world (BW) \cite{Randall} model is based on the assumption that
our four dimensional space-time is a three-brane, embedded in a five dimensional bulk.
Braneworld models have an appreciable impact in theoretical physics in so far as offering solutions to the mass hierarchy problem in particle physics. The Randall-Sundrum
(RS) braneworld models play a significant role in cosmology, in particular, these models provide an explanation for the expansion rate
of the universe at high energies which differs from the prediction of standard general relativity \cite{Hawkins}.

Exact solutions describing braneworld stellar models are few and far in between. Germani and Maartens presented a static spherically
star solution with uniform density\cite{Germani}. An elegant overview of stars on the brane which include the matching conditions and projection of the Weyl stresses from the bulk was presented by Deruelle\cite{D}.
Some elegant work on star solutions in the braneworld scenario have been discussed in \cite{Ovalle1,Ovalle2,Linares(2015),Francisco}. Gravitational collapse of bounded matter configurations on the brane was discussed by Govender and Dadhich\cite{govender}.
Conformal symmetries are of some importance for better understanding of spacetime geometry
because  it helps solve the geodesic equations of motion for the spacetime under consideration.
Symmetry also helps to search for natural relationship between geometry and matter.

 In relativity theory, the behavior of the metric is important when moved along curves on a manifold. Conformal killing vector $\xi$ is a vector field on a manifold so that if the metric is dragged along the curves created by $\xi$,  its Lie derivative is directly proportional  to itself i.e.
 \begin{equation}
\mathcal{L}_\xi  g_{ik}=\psi  g_{ik},
\end{equation}   for some scalar field $ \psi$, known as conformal factor.
  Here $\mathcal{L}$ represents the Lie derivative operator .
The physical importance of this  prerequisite is that when the metric is dragged along specific congruence of curves it    persists itself modulo some scale factor, $ \psi$, which may   differ  from  position  to position on the manifold. One can note that $ \psi$ is not arbitrary but it depends on the conformal killing vector $ \xi$ as
$\psi(x^k) = \frac{1}{4} \xi_{:i}^i $ for Riemannian space of dimension four.

%A Conformal Killing Vector (CKV) satisfies
%\begin{equation}
%\mathcal{L}_\xi g_{ik}=\psi g_{ik},
%\end{equation}
%where $\mathcal{L}$ represents the Lie derivative operator and $ \psi$ the conformal
%factor, respectively with $\xi_i= g_{ik}\xi^k$.
 The vector $ \xi$  characterizes the conformal symmetry,
while the metric tensor $g_{ik}$ is conformally mapped onto
itself along $\xi$. The application of CKV provides
deeper insights into the spacetime geometry. The  conformally symmetric vacuum solutions
of the gravitational field equations in the brane-world models have been found in \cite{Harko}.
Usmani et al., found the gravastar solution within the framework of the
Mazur-Mottola, admitting conformal motion in \cite{Usmani}.

In this paper we present the solutions for gravastar in the context of Randall-Sundrum II
type braneworld scenario. The paper is organized as follows: in section \textbf{II} and \textbf{III} we briefly
review the field equations in brane world models by admitting conformal motion of Killing Vectors.
In section \textbf{IV} we derive the stellar interior solution based on the fact that the matter in the core obeys a barotropic equation of state of the form $p = -\rho > 0$. The Weyl stresses, namely the scalar $\mathcal{U}$ and the anisotropy $\mathcal{P}$, in this region is completely determined.
In section \textbf{V} we present the complete gravitational and thermodynamical behaviour of the shell of the star while the external region has been matched
to the braneworld black hole solution in section \textbf{VI}. In section \textbf{VII} we calculate
the entropy of the fluid within the shell and present the junction conditions for the solutions
under consideration in section \textbf{VII}. Some final remarks are made
in the Section \textbf{IX}.

\section{ Field Equations}
Let us start by writing the modified Einstein field equations on the brane \cite{Shiromizu}, which take
the form
\begin{equation}
G_{\mu\nu}=k^2 T^{\text{eff}}_{\mu\nu},
\end{equation}
where $k^2=8\pi G$ and $T^{\text{eff}}_{\mu\nu}$ represent the effective energy-momentum tensor
given by
\begin{equation}
 T^{\text{eff}}_{\mu\nu}= T_{\mu\nu}+\frac{6}{\lambda}S_{\mu\nu}-\frac{1}{k^2}\mathcal{E}_{\mu\nu},
\end{equation}
We have chosen the bulk cosmological constant in such a way that the brane cosmological
constant vanishes and $\lambda$ is the brane tension which corresponds to the vacuum energy density on the brane. The high-energy and non-local corrections respectively are given by
\begin{equation}
S_{\mu\nu}=\frac{TT_{\mu\nu}}{12}-\frac{T_{\mu\alpha}T^{\alpha}_{\nu}}{4}+\frac{g_{\mu\nu}}{24}
\left[3T_{\alpha\beta}T^{\alpha\beta}-T^2\right],
\end{equation}
where $T^{\alpha}_{\alpha}$, and
\begin{equation}
k^2 \mathcal{E}_{\mu\nu}=-\frac{6}{\lambda}\left[\mathcal{U}\left(u_{\mu}u_{\nu}+\frac{1}{3}h_{\mu\nu}\right)
+\mathcal{P_{\mu\nu}}+\mathcal{Q}_{(\mu u_{\nu})}\right],
\end{equation}
is a non-local source, arising from the 5-dimensional Weyl curvature with $\mathcal{U}$
representing the nonlocal energy density, $\mathcal{Q}_{\mu }$ is the nonlocal energy
flux and nonlocal  anisotropic pressure is $ \mathcal{P}_{\mu\nu }$, respectively.

 For a static spherically symmetric matter distribution $\mathcal{Q}_{\mu } = 0 $ and the
 nonlocal anisotropic pressure  $ \mathcal{P}_{\mu\nu }$ is given by
\begin{equation}
\mathcal{P}_{\mu\nu }=\mathcal{P}\left(r_{\mu }r_{\nu}-\frac{1}{3}h_{\mu\nu } \right),
\end{equation}
where $r_{\mu }$ is the projected radial vector and $h_{\mu \nu}= g_{\mu \nu}+ u_{\mu }u_{\nu}$ is the
projected tensor with the  4-velocity $u^{\mu}$ and P is the  pressure  of the bulk. The above expression  becomes
\begin{equation}
k^2 \mathcal{E}_{\mu\nu}=-\frac{6}{\lambda}\left[\mathcal{U}u_{\mu}u_{\nu}+\mathcal{P}r_{\mu}r_{\nu}
+h_{\mu\nu}\frac{(\mathcal{U}-\mathcal{P})}{3}\right],
\end{equation}
where $r_{\mu}$ is a unit radial vector. We consider a perfect fluid energy-momentum
tensor $T_{\mu\nu}$, having the explicit form
\begin{equation}
T_{\mu\nu}=\rho u_{\mu}u_{\nu}+p h_{\mu\nu},
\end{equation}
where $u^{\mu}$ is the 4-velocity. We consider the static spherically symmetric line
element on the brane in the standard form
\begin{equation} \label{metric}
ds^2=- e^{\nu(r)}dt^2+ e^{\lambda(r)}dr^2+r^2\left(d\theta^{2}+\sin^{2}\theta d\phi^2\right).
\end{equation}
The gravitational field equations for the line element metric (\ref{metric}) must satisfy the effective 4-D Einstein
equations (1) with the effective energy-momentum tensor \cite{Maarteens,Koyama,Ovalle}
\begin{equation}
e^{-\lambda}\left[\frac{\lambda'}{r}-\frac{1}{r^{2}} \right]+\frac{1}{r^{2}}=k^2 \rho^{\text{eff}},
\end{equation}
\begin{equation}
e^{-\lambda}\left[\frac{1}{r^{2}}+\frac{\nu'}{r} \right]-\frac{1}{r^{2}}=k^2\left(p^{\text{eff}}+\frac{4}{k^{4}\lambda }\mathcal{P}\right),
\end{equation}
\begin{equation}
\frac{1}{2}e^{-\lambda}\left[\frac{1}{2}\nu'^{2}+\nu''-\frac{1}{2}\lambda{'}\nu'+\frac{1}{r}(\nu'-\lambda{'}) \right]=k^2 \left(p^{\text{eff}}-\frac{2}{k^{4}\lambda }\mathcal{P}\right),
\end{equation}
\begin{equation}
p^{\prime}+\frac{\nu^{\prime}}{2}\left(\rho+p\right)=0,
\end{equation}
where primes denote the differentiation with respect to $r$ and
 the effective energy density $\rho^{\text{eff}}$, effective radial pressure $\tilde{p}_r$ and
 effective transverse pressure $\tilde{p}_t$ are given by
\begin{equation} \label{eff}
\rho^{\text{eff}}=\tilde{\rho}=\rho\left(1+\frac{\rho}{2\lambda}\right)+\frac{6}{k^4 \lambda}\mathcal{U},
\end{equation}
\begin{equation}
p^{\text{eff}}_r=\tilde{p}_r=p+\frac{\rho}{2\lambda}\left(\rho+2p \right)+\frac{2}{k^4 \lambda}\mathcal{U}+\frac{4}{k^4 \lambda}\mathcal{P},
\end{equation}
\begin{equation}
p^{\text{eff}}_t=\tilde{p}_t=p+\frac{\rho}{2\lambda}\left(\rho+2p \right)+\frac{2}{k^4 \lambda}\mathcal{U}-\frac{2}{k^4 \lambda}\mathcal{P}.
\end{equation}
Moreover, we observe from Eq. (5), that $\mathcal{E}_{\mu\nu}$$\rightarrow 0$ as $\lambda^{-1} \rightarrow 0$,
i.e., using the limit in Eq. (3), we obtain $T ^{\text{eff}}_{\mu\nu} = T _{\mu\nu}$,
thereby recovering 4-dimensional general relativity.
The extra dimensional effects produce anisotropy in the interior of the star distribution which
can be written as
\begin{equation}
\tilde{p}_r-\tilde{p}_t=\frac{6}{k^4 \lambda}\mathcal{P}.
\end{equation}
Recently, Mazur and Mottola  \cite{MM2015} made an interesting observation that the constant density interior Schwarzschild solution for a static, spherically symmetric collapsed star has a negative  pressure when its radius is  less than Schwarzschild radius  thereby describing a gravitational condensate star or gravastar. Furthermore, they showed that transverse stresses are induced within this region thereby abandoning the condition of pressure isotropy. In a more general approach Cattoen et al. \cite{cattoen} showed that stable gravastars must necessarily exhibit transverse pressures. It is interesting to note from (17) that our braneword construction of a gravastar naturally incorporates an effective transverse pressure arising from nonlocal Weyl stresses.

\section{Gravastar with conformal motion }

We now demand that the interior spacetime admits conformal motion. This immediately places a restriction on the gravitational behaviour of the gravastar. From Eq. (1) we can write
\begin{equation}
\mathcal{L}_\xi g_{ik}=\xi_{i;k}+\xi_{k;i}=\psi g_{ik},
\end{equation}
with $ \xi_{i}=g_{ik}\xi^{k}$ .
From Eqs. $(9)$ and $(18)$, one can get the following expressions
$ \xi^{1}\nu'=\psi $ ,
$ \xi^{4}= C_1$ ,
$ \xi^{1}= \frac{\psi r}{2}$,
$ \xi^{1}\lambda' + 2\xi_{,1}^{1}=\psi$ .\\

Here, $1$ and $4$ represent the spatial and temporal coordinates $r$ and $t$ respectively.
The above set of equations imply
\begin{eqnarray}
e^{\nu}&=&C_{2}^{2}r^{2},\\
e^{\lambda}&=&\left[\frac{C_3}{\psi}\right]^{2},\\
\xi^{i}&=&C_{1}\delta_{4}^{i}+ \left[\frac{\psi r}{2}\right]\delta_{1}^{i},
\end{eqnarray}
where  $C_{1}$, $C_2$, and $C_{3}$ are constants of integration.
The field equations (Eq. 10-12) corresponding to the metric $(9)$ take the following
form
\begin{eqnarray} \label{dark}
&&\frac{1}{r^2}\left[1-\frac{\psi ^2}{C_{3}^2}\right]-\frac{2\psi \psi^{\prime}}{rC_{3}^2}=k^2 \rho^{\text{eff}} \label{dark1},
\\ \nonumber \\
&&\frac{1}{r^2}\left[\frac{3\psi ^2}{C_{3}^2}-1\right]=k^2 \left(p^{\text{eff}}+\frac{4}{k^{4}\lambda }\mathcal{P}\right) \label{dark2},
\\ \nonumber \\
&&\left[\frac{\psi ^2}{C_{3}^2 r^2}\right]+\frac{2\psi \psi^{\prime}}{rC_{3}^2}
=k^2 \left(p^{\text{eff}}-\frac{2}{k^{4}\lambda }\mathcal{P}\right) \label{dark3},
\end{eqnarray}
The above system of equations represents a spherically symmetric matter distribution admitting conformal motion on the brane.
From Eq. (\ref{dark2}) and Eq. (\ref{dark3}) we can express the extra dimensional effects in terms of
 conformal factor as
\begin{equation} \label{extrap}
 \frac{6}{k^2 \lambda }\mathcal{P}=\frac{2\psi ^2}{C_3^2 r^2}-\frac{1}{r^2}-\frac{2\psi  \psi ^{\prime}}{C_3^2 r},
\end{equation}
 In order to obtain the complete gravitational behaviour of the model, the function $\psi$ needs to be determined.

\section{Interior region of the Braneworld Gravastar }

We now consider the inner portion of our gravastar model. When p = $-\rho$, Eq. (13) yields $\rho= {\text const.} = \rho_c$,
therefore $ p = p_{c} = -\rho_c$.
Using (\ref{extrap}) in (\ref{dark2}) and  together with the equation of state we obtain
\begin{equation}
 \frac{2}{k^2 \lambda }\mathcal{U}=-\frac{1}{3r^2}+\frac{5\psi ^2}{3C_3^2 r^2}+\frac{4\psi \psi^{\prime}}{3C_3^2 r}
 +k^2\rho_c\left(1+\frac{\rho_c}{2\lambda}\right).
\end{equation}
Now using the values of  $\mathcal{U}$ and $\mathcal{P}$ in Eq.(22) we obtain

\begin{equation} \label{psi}
 \psi ^2+ r \psi \psi ^{\prime}= \frac{C_3^2}{3}\left(1-2\mathcal{A}r^2\right),
\end{equation}
where $\mathcal{A}$= $k^2\rho_c\left(1+\frac{\rho_c}{2\lambda}\right)$. Solving (\ref{psi}) we obtain
\begin{equation}
 \psi ^2= \psi_0+\frac{C_3^2}{3}\left(1-\mathcal{A}r^2\right),
\end{equation}
where  $\psi_0$ is an integration constant. Now, the metric potential $e^\lambda$ assumes the following form as
\begin{equation}
e^{\lambda} =  C_3^2\left[\psi_0+\frac{C_3^2}{3}\left(1-\mathcal{A}r^2\right)\right]^{-1}.
\end{equation}

 The  radiation energy density  and   pressure   of the bulk are obtained as respectively,
\begin{equation}
 \frac{2}{k^2 \lambda }\mathcal{U}=\frac{1}{3 r^2}\left(\frac{2}{3}+\frac{5\psi_0}{C_3^2 r}\right),
\end{equation}
\begin{equation}
 \frac{6}{k^4 \lambda }\mathcal{P}=\frac{1}{3 r^2}\left(\frac{6\psi_0}{C_3^2}-1\right),
\end{equation}
For real conformal factor, it is clear from eq. (28) that
$r< \sqrt{\left[\frac{3\psi_0}{\mathcal{A}C_3^2}+\mathcal{A}\right]}$. This gives a clue of the upper limit
of the interior region.
 Since   the conformal factor $ \psi$  may  vary   from  place  to place  on the manifold, this  implies that it places a restriction on the size of the interior region.

 For $\psi_0 = 0$, one can find that both $\mathcal{U}$
 and $\mathcal{P}$ are inversely proportional to $r^2$ but with opposite signs.
  Note that for $\psi_0 > \frac{C_3^2}{6}$ both $\mathcal{U}$ and $\mathcal{P}$ are positive, however,
  for $\psi_0 < \frac{C_3^2}{6}$, $\mathcal{U}$ takes positive value but $\mathcal{P}$ has negative value.
  For for $\psi_0 = \frac{C_3^2}{6}$, radiation pressure of the bulk vanishes and only
  radiation energy density $\mathcal{U}$ of the bulk survive.

The active gravitational mass $M(r)$, can be obtained from Eq. (\ref{dark1}), which may be expressed
in the form
\begin{equation}
M(r)= 4\pi \int_0^r  \rho^{\text{eff}} \tilde{r}^2  d\tilde{r}
=\frac{1}{2G}r\left[1-\left(\frac{\psi_0}{C_3^2}+\frac{1}{3}
\left(1-\mathcal{A}r^2\right)\right)\right].
\end{equation}
The radiation energy density and pressure of the bulk fail to be regular at the origin, however, the effective gravitational mass is always positive and vanishes as $r \rightarrow 0$. This implies that the effective gravitational mass is singularity free.

\section{Shell of the gravastar }

For the shell of the gravastar we consider a
ultra-relativistic fluid of soft quanta obeying the EOS $p = \rho$, which
represents a stiff fluid.  This EOS is referred to as the Zel'dovich Universe which has
been studied by various authors for in the cosmological context\cite{Zel'dovich,Carr,Madsen} and
astrophysical settings \cite{Braje,Linares,Wesson}.

Now, using the EOS $p=\rho$, we obtain  from Eq. (13)
\begin{equation}
p= \rho= \frac{m}{r^2},
\end{equation}
where $m$ is an integration constant. Using the same
expression given in (\ref{extrap}), we are in a position to determine $\mathcal{U}$, from Eq. (23), which is given by
\begin{equation}
 \frac{2}{k^2 \lambda }\mathcal{U}=-\frac{1}{3r^2}+\frac{5\psi ^2}{3C_3^2 r^2}+\frac{4\psi \psi^{\prime}}{3C_3^2 r}
 -\frac{mk^2}{r^2}-\frac{3k^2m^2}{2\lambda r^4}.
\end{equation}
Using the values of  $\mathcal{U}$ and $\mathcal{P}$ in Eq. (22) we obtain
\begin{equation}
 \psi ^2+ r \psi \psi ^{\prime}= \frac{C_3^2}{3}\left[(1+mk^2)+\frac{2k^2 m^2}{\lambda r^2}\right],
\end{equation}
which is easily solved to yield
\begin{equation}
 \psi ^2= \psi_1+\frac{C_3^2}{3}\left[(1+mk^2)+\frac{4k^2m^2}{\lambda}\frac{\ln r}{r^2}\right].
\end{equation}
[ $\psi_1$ is an integration constant ]

This allows us to write
\begin{equation}
\mathcal{U}=\frac{k^2 \lambda \psi_1}{ 6C_3^2r}\left(\frac{5}{r}-4\right)+ \frac{k^2 \lambda}{9 r^2}
 \left(1-2k^2 m\right)+\frac{k^4  m^2}{9r^4}\left(2\ln r-\frac{11}{4}\right),
\end{equation}
and
\begin{equation}
 \mathcal{P}=\frac{k^2 \lambda \psi_1}{ 3C_3^2r}\left(1+\frac{1}{r}\right) -\frac{k^2 \lambda}{18r^2}+
 \frac{ mk^4}{9 r^2}\left(\lambda-\frac{2m}{r^2}+8m\frac{\ln r}{r^4}\right).
\end{equation}
Here we assume the interfaces at $r = R$ and $r = R+\epsilon$ which describe the matching
surface of two regions i.e., interior and exterior region, are very thin. This means that
$\epsilon$ is very small i.e., 0$<\epsilon$ $\ll 1$. The proper thickness between two interfaces
is obtained as:
\begin{equation}
l=\int^{R+\epsilon}_{R} \sqrt{e^{\lambda}} dr =\int^{R+\epsilon}_{R}\frac {1} {\sqrt{f(r)}} dr.
\end{equation}
Let we introduce $H$ as the primitive of $\frac {1} {\sqrt{f(r)}}$ i.e., $\frac {dH}{dr}=\frac {1} {\sqrt{f(r)}}$,
which provides the proper thickness as:
\begin{equation}
l=[H]^{R+\epsilon}_{R}.
\end{equation}
Now, using the Taylor series expansion $H(R+\epsilon$) up to first order approximation about $R$, we obtain
$H(R+\epsilon$)$\simeq$ H(R)+$\epsilon$ $H^{\prime}$(R), and we can write
$l$ $\approx$ $\epsilon \frac {dH}{dr}$ $|_{R} $.
Therefore using the Eq. (39), the proper thickness is finally given by
\begin{equation}
l \approx \epsilon \left( \frac{C_3^2}{\psi_1}+\frac{3}{(1+mk^2)+\frac{4k^2m^2}{\lambda}\frac{\ln R}{R^2}}\right)^{1/2}.
\end{equation}
Now, the energy $\mathcal{E}$, within the shell is given by
\begin{equation}
\mathcal{E}= 4\pi \int^{R+\epsilon }_{R} \rho^{eff}~ r^2 dr=\left[ 4 \pi m \left(r-\frac{m}{2 \lambda r}\right)+ \frac{4 \pi \psi_1}{k^2C_3^2}(5 r-2r^2)
 + \frac{8 \pi r}{3k^2}\left(1-2k^2m\right) +\frac{8 \pi m^2}{9 \lambda }\left(- \frac{\ln r}{r}+\frac{7}{4r}\right)\right]^{R+\epsilon }_{R}.
\end{equation}
It is to be noted that the energy within the shell  depends on the thickness of the shell as well as brane tension.

\section{Exterior region of the gravastar }

In general the exterior spacetime on the brane is nonempty, ie., $\rho^{eff}$ and $p^{eff}$ are nonzero. This is due to the presence of the Weyl stresses brought about by bulk graviton effects. In order to close the system of equations in the exterior region one has to prescribe ${\cal P}^{+}$ and ${\cal U}^{+}$ which are not unique. The simplest choice ${\cal P}^{+} = {\cal U}^{+} = 0$ ensures that the exterior is the vacuum Schwarzschild solution. For non vanishing Weyl stresses in the exterior $(p= \rho =0)$, the spacetime is described by the metric
\begin{equation}
ds^{2}=-f(r)dt^2+f^{-1}(r)dr^2+r^2\left(d\theta^2+\sin^2 \theta d\phi^{2}\right),
\end{equation}
where $f(r)= 1-\frac{2M}{r}+\frac{q}{r^2}$.

This exterior   spacetime  corresponds to
Tidal Charged Black hole   characterized by two parameters:  their mass M and dimension less
tidal charge, q.  Here, the  tidal charge parameter q    emanates from the prognosis on the brane
of free gravitational field effects  in the bulk.  This    tidal charge  q may assume
positive and negative values. For positive  tidal charge   the metric
(43) corresponds to Reissner-Nordstr${\ddot{o}}$m Black hole solution.     The presence of the tidal charge increases the  gravitational field
of this black hole.

\section{Entropy within the shell}

We shall try to calculate the entropy of the fluid within the shell by adapting the concept
of Mazur and Mottola \cite{Mazur}, which is given by
\begin{equation}
\mathcal{S}= 4\pi\int^{R+\epsilon}_{R} sr^2\sqrt{e^{\lambda}}~ dr =4\pi \epsilon s r^2\sqrt{e^{\lambda}}.
\end{equation}
Here, s(r) is the entropy density for the local temperature T(r), which may be
written as
\begin{equation}
s(r)=\frac{\alpha^2 k_B^2 T(r)}{4\pi \hbar^2 G} =\alpha \left(\frac{k_B}{\hbar}\right)\sqrt{\frac{p}{2\pi G}},
\end{equation}
where $\alpha^{2}$ is a dimensionless constant.\\
Thus the entropy of the fluid within the shell is (applying the same procedure to obtain Eq. (39)), given by
\begin{equation}
\mathcal{S}=\sqrt{\frac{m}{2\pi G}}\left(\frac{4\pi \epsilon \alpha k_B R}{\hbar}\right) \left( \frac{C_3^2}{\psi_1}+\frac{3}{(1+mk^2)+\frac{4k^2m^2}{\lambda}\frac{\ln R}{R^2}}\right)^{1/2}.
\end{equation}

\section{Junction interface and surface stresses}
Here we match the interior gravastar geometry, given in Eq. (9), with an
exterior Braneworld black hole solution at a junction interface $\Sigma$. The
junction surface $\Sigma$, is a timelike hypersurface defined by the parametric
equation $f\left(x^\mu(\xi^i)\right)=0$, where $\xi^i= (\tau, \theta, \phi)$
are the intrinsic coordinates on the hypersurface with proper time $\tau$.

The extrinsic curvature (second fundamental form) associated with a hypersurface $\Sigma$ is given by
\begin{equation}
K^{\pm}_{ij} = -\eta_{\nu}\left(\frac{\partial^2 x^{\nu}}{\partial\xi^i \partial\xi^j}+\Gamma^{\nu\pm}_{\alpha\beta}\frac{\partial x^{\alpha}}{\partial\xi^{i}}\frac{\partial x^{\beta}}{\partial\xi^{j}}\right),
\end{equation}
where $\eta_{\nu}$ represents the unit normal $(\eta^{\nu}\eta_{\nu}=1)$ at the
junction defined by
\begin{equation}
n_{\nu} = \pm \Bigg \vert g^{\alpha\beta}\frac{\partial f}{\partial x^{\alpha}}\frac{\partial f}{\partial x^{\beta}} \Bigg \vert ^{-1/2}\frac{\partial f}{\partial x^{\nu}}.
\end{equation}
Since the $K_{ij}$ is discontinuous across the $\Sigma$ and the discontinuity of the metric
is usually described by $k_{ij}= K^{+}_{ij}-K^{-}_{ij}$.

Now using the Lanczos equation \cite{Israel,Rahaman (2009),Rahaman ((53)2009)}, we can
obtain the intrinsic stress-energy tensor as $S^{i}_j$ = diag ($-\sigma, \mathfrak{p}, \mathfrak{p}$),
where $ \sigma$ is the line energy density  and $\mathfrak{p}$ is the line tension
and defined by
\begin{equation}
\sigma=-\frac{1}{4\pi a}\left[\sqrt{e^{-\lambda}}\right]^{+}_{-} \\
= -\frac{1}{4\pi a}\left[\sqrt{1-\frac{2M}{a}+\frac{q}{a^2}}
-\sqrt{\frac{\psi_0} {C_3^2}+\frac{1}{3}\left(1-\mathcal{A}a^2\right)}\right].
\end{equation}
\begin{equation}
\mathfrak{p}=\frac{1}{8\pi a}\left[\left(1+\frac{a\nu^{\prime}}{a}\right)\sqrt{e^{-\lambda}}\right]^{+}_{-} \\
= \frac{1}{4\pi a}\left[\frac{1-M/a}{2\sqrt{1-\frac{2M}{a}+\frac{q}{a^2}}}-\sqrt{\frac{\psi_0} {C_3^2}+\frac{1}{3}\left(1-\mathcal{A}a^2\right)}\right].
\end{equation}

The surface mass of the thin shell is defined by
\begin{equation}
m_s= 4\pi a^2 \sigma= -a\left[\sqrt{1-\frac{2M}{a}+\frac{q}{a^2}}
-\sqrt{\frac{\psi_0} {C_3^2}+\frac{1}{3}\left(1-\mathcal{A}a^2\right)}\right].
\end{equation}
Here $M$ can be interpreted as the total mass of the braneworld black hole solution. It can be written
in the following form:
\begin{equation}
M= \frac{1}{2a}\left[a^2+q-a^2\zeta^2+2am_s\zeta-m_s^2 \right],
\end{equation}
where $\zeta^2 (a)=\frac{\psi_0} {C_3^2}+\frac{1}{3}\left(1-\mathcal{A}a^2\right)$.
For an interesting observation we note that the line tension is negative which implies that there is a line
pressure as opposed to a line tension. In our configuration, the junction interference i.e. region $(b)$
contains two different types of fluid: one is ultra-relativistic fluid obeying p = $\rho$, and the other
is matter component which arises from the discontinuity of the second fundamental form. This provides extra
surface stress energy and surface tension at the junction interface.

\section{Final Remarks}

In our work we have focused our attention on the problem of modeling gravastars within the context of Randall-Sundrum II type braneworld scenario, based on the fact that our four dimensional  space-time is a three-brane, embedded in a five dimensional bulk.
The star model has three distinct regions with different equations of state:(a) interior
solution: $0 \leq r \leq r_1$, with equation of state $p = -\rho > 0$, (b) shell: $r_1 < r < r_2$ with equation
of state $p = +\rho$ and (c) exterior: $r_2 < r$ with $p = \rho = 0$ and solution is found
under the assumption of conformal motion.

In our model, we have shown that the upper limit of the interior region
can be derived for real conformal factor (i.e., $\psi > 0$). For an interesting observation
the dimensionless integration constant $\psi_0$, plays an important role in determining
the nature of the bulk Weyl scalar $\mathcal{U}$ and nonlocal anisotropic pressure $\mathcal{P}$.
At $\psi_0 = C_3^2/3$ represents a barrier, where the radiation pressure of the bulk vanishes but
effect of bulk Weyl scalar is nonzero and takes positive values at greater radii. In the core region
both $\mathcal{U}$ and $\mathcal{P}$ fail to be regular at the origin, however the effective gravitational
mass is always positive and vanishes at $r = 0$, which provides a singularity free solution for our
model. Next we have discussed the explicit form of the shell of the star by considering an ultra-relativistic
fluid of soft quanta obeying the EOS $p= \rho$, and found the proper thickness of the shell. We also explicitly show that the energy
within the shell depends on the thickness of the shell as well as brane tension.
We have also shown that the junction interference
contains two different types of matter, namely ultra-relativistic fluid and matter components appear due to the discontinuity of the affine connections at the region b.
The later one is a thin shell of
matter content with negative energy density.   This newly developed  stress-energy tensor supports   the
consideration of Casimir effect between compact objects at
arbitrary separations \cite{Emig2007}. It is argued that   these two fluids  do not interact  and characterize the shell of the gravastar.
It is  shown that the junction interference  between interior and exterior regions contains   a thin shell of matter   with negative energy density.  It seems that this negative energy  is very similar to the Casimir effect reported in reference  \cite{Emig2007} in which the authors have obtained  the electromagnetic  Casimir interaction between compact objects at arbitrary separations.
We conclude with a vital
point: this paper is not intended to confirm exact alternative to braneworld black hole, but to make us confirm that this is
the first attempt of a project to propose the braneworld  gravastar configuration which is
alternative to braneworld black hole. We constructed a model of a braneworld gravastar by requiring that the interior spacetime admits conformal motion. We have  integrated  the resulting equations and obtained  a class of exact solutions that describe gravastars on the brane which are particular alternatives to the braneworld black hole.
Now, it is an obvious query whether  this braneworld construction change in any way the implications of a gravastar solution
to the black hole paradoxes already proposed without any braneworld?
  It needs further research. Hopefully, it will be a subject for future study. However, we must point out that the braneworld construction of a gravastar naturally incorporates anisotropic pressure via the brane embedding. This may not be the case in standard general relativity, that is to say,  the anisotropy required for stability of gravastars needs to be assimilated into the energy momentum tensor a priori.

\subsection*{Acknowledgments}
AB and FR would like to thank the authorities of the Inter-University Centre
for Astronomy and Astrophysics, Pune, India for providing the Visiting
Associateship under which a part of this work was carried out. FR and SI are also
thankful to DST, Govt. of India for providing financial support
under PURSE programme and INSPIRE Fellowship respectively.  We are grateful to the referee for his valuable suggestions. We are also grateful to Prof. G S Khadekar for helpful discussion.

\end{document}